# Emergence of coherent backscattering from sparse and finite disordered media

NOOSHIN M. ESTAKHRI,[1, *] NASIM MOHAMMADI ESTAKHRI,[2] AND THEODORE B. NORRIS[3]

[1]*Department of Physics, Virginia Tech, Blacksburg, Virginia 24061, USA*
[2]*Fowler School of Engineering, Chapman University, Orange, CA 92866, USA*
[3]*Department of Electrical Engineering and Computer Science, University of Michigan, Ann Arbor, Michigan 48109, USA*
*\*estakhri@umich.edu*

**Abstract:** Coherent backscattering (CBS) arises from complex interactions of a coherent beam with randomly positioned particles, which has been typically studied in media with large numbers of scatterers and high opacity. We develop a first-principles scattering model for scalar waves to study the CBS cone formation in finite-sized and sparse random media with specific geometries. The results provide new insights into the effects of density, volume size, and other relevant parameters on the angular characteristics of the CBS cone emerging from bounded random media for various types of illumination. This work also highlights some of the potentials and limitations of employing the coherent backscattering phenomenon to characterize disordered configurations. The method developed here provides a foundation for studies of the multiple scattering of complex electromagnetic fields in randomized geometries, including quantized fields for investigating the effects of the quantum nature of light in multiple scattering.

## 1. Introduction

The phenomenon of coherent backscattering (CBS) is an expression of the weak localization of waves in disordered media and stems from coherent effects among time-reversed scattering paths, which introduce identical phases in multiple scattering [1, 2]. The constructive interference between reciprocal scattering paths [3], leads to an enhancement in the average intensity of the scattered light in the backscattering direction. While scattering from any specific realization of disorder leads to speckle, CBS may be observed by ensemble averaging over different realizations, ideally exhibiting a two-fold enhancement of the intensity in the exact backscattering direction compared to the diffuse background at large angles. Enhanced backscattering was first considered in studies of the electromagnetic reflection from a turbulent medium, looking at the large-scale interference effects [4]. The CBS effect has since been observed for multiple scattering of electromagnetic waves in several experiments including studies on concentrated aqueous suspension of dielectric spheres [5, 6, 7], dielectric powders [8, 9], multimode optical fibers [10], photonic crystals [11], cold atoms [12, 13], and biological media [14, 15]. In addition, the formation and behavior of this coherent effect have been studied in the presence of nonlinearity [16, 17], time variation [18], and magnetic biases [10, 19, 20]. In principle, reciprocity must be maintained for CBS to appear, and any perturbation inducing nonreciprocal behavior may alter or eliminate such localization effects [21]. For instance, nonlinearity introduces additional phase shifts proportional to the wave intensity, which can transform CBS peak into a dip (dephasing and antilocalization). Additionally, CBS peaks have been studied and observed for various types of waves besides electromagnetic waves [22, 23, 24, 25, 26].

A wide variety of theoretical approaches have been taken to model light scattering in random media, employing microscopic or macroscopic descriptions of the scattering phenomena. Multiple scattering theory for discrete scatterers has been employed using Feynman diagrams. The backscattering enhancement, however, cannot be captured by using ladder terms in such analyses and it is necessary to use cyclical terms, as demonstrated in [3]. Inclusion of all ladder terms and cyclical terms becomes necessary if the optical thickness of the random medium becomes appreciable and for samples with high surface reflectivity [27]. Classical transport theory has also been used to estimate the Green's function describing light transport in a random medium, following the diffusion approximation [1]. This technique has allowed for the estimation of the shape and width of the backscattering cone [1]. Furthermore, numerical simulations such as Monte Carlo and related approaches [18, 28, 29, 30], and T-Matrix method [31, 32] have also been used to model the multiple scattering processes in disordered media.

In this paper, we use a microscopic numerical scalar wave analysis technique to examine the formation of CBS peaks in scattering from finite-sized, wavelength-scale random media. The inherent simplicity of the applied procedure allows us to systematically study how different parameters of the scattering object can affect the line shape of the backscattered enhancement, and specifically to understand the limits on the formation of the backscattering cone for small numbers of scatterers and sparse samples. Additionally, this approach allows a straightforward way to calculate the effects of density factor, sample size, absolute number of particles, and iteration factor. Here, we first consider scattering from small scatterers positioned within a rectangular space with different edge sizes ranging from 10 to 40 free-space wavelengths, with excitation in the form of a monochromatic plane wave. This approach also allows us to compute the scattering of arbitrary input states of light, such finite-size excitation beams or other tailored excitations [14, 33]. In addition to the emergence of CBS, we analyze the ensemble-averaged specular reflection in the scattering from random samples, a feature typically overlooked in previous numerical studies. We then develop a method to eliminate this coherent specular contribution, and hence make it possible to observe the pure CBS peak under normally incident illumination.

While a scalar analysis does not capture polarization effects [34, 35, 36] (e.g., the polarization opposition effect [34]), it allows us to study more populated scattering volumes. Here, we analyze volumes consisting of up to 20000 particles, much higher than previous studies which calculated the exact solutions of Maxwell's equations [37].

## 2. Formalism and modeling

To model the CBS formation, a random medium is typically approximated by a half space in order to simplify the calculation [1,3,38]. This approximation is accurate for cases of centimeter-sized samples under external optical illumination, as the mean free path length of the wave inside such disordered media is much smaller than the physical size of the scattering sample. However, such models do not provide insights into the behavior and physics of wave scattering in the case of smaller random configurations, i.e., those with dimensions in the order of multiple wavelengths or comparable to the mean free path length, as we study in this work.

Figure 1 depicts a sketch of the finite-sized random media examined in this study together with the direction of illumination with respect to the chosen coordinate system. A total of $N_p$ isotropic small particles (scattering centers) are assembled inside an imaginary rectangular domain, where the center of each particle is randomly positioned. We use uniform statistical distributions to assign all three components of the particles' positions (i.e., x, y, and z); and the process is repeated $N_r$ times to realize $N_r$ independent random distributions. In experiments, ensemble averaging of the scattered intensity, which results in a significant reduction of granular behavior (speckle), is typically achieved through rotating or moving the sample. This step is executed here, however, by computing fully independent realizations of the composition followed by ensemble averaging over the calculated results. The uniform distribution of the particles' positions (which could be altered to any other distribution) is the only assumption made about the scattering medium, and we do not impose any assumptions on the statistical behavior of the response or the scattering matrix elements. It is noteworthy that since $N_p$ specifies the number of equations that needs to be solved at each iteration, changing the distribution of particles, their composition, or the domain size, does not affect the computation load and processing time. We have found that in this model, a typical personal computer with 32GB RAM is capable of handling $N_p$ values in the order of up to about $10^4$; no cloud computing or advanced computing techniques were required to obtain the results presented here. The structure is illuminated with a scalar electromagnetic plane wave propagating in the xz-plane, incident upon the structure at the angle $\theta_i$, as shown in Fig. 1. The far field scattering pattern from the particle cluster for all outgoing wavevectors (depicted as $\theta_s$) is then calculated. For $-\pi/2 \leq \theta_s \leq \pi/2$ the entire upper half-plane region is covered, in which $\theta_s = -\theta_i$ corresponds to the backscattering direction and $\theta_s = \theta_i$ corresponds to the specular reflection of the incident wave. With the finite size of the sample, the random object also creates scattering in the forward direction (i.e., $|\theta_s| \geq \pi/2$), which is not of interest in our studies. In the back half-plane, as will be discussed below, the observed coherent effects in multiple scattering, which survive the ensemble averaging, are mainly at the specular ($\theta_s = \theta_i$) and the backscattering ($\theta_s = -\theta_i$) directions.

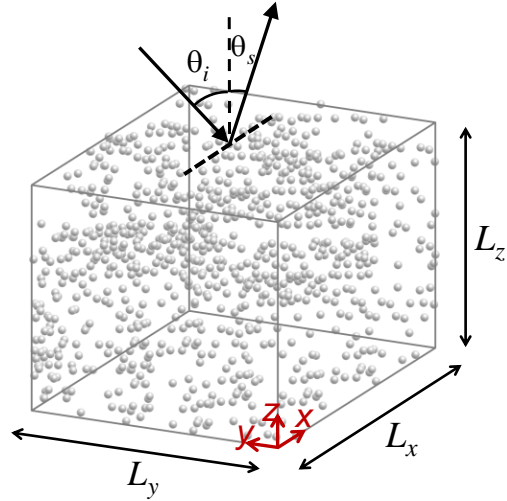

Fig. 1. Simulation setup composed of $N_p$ particles randomly arranged inside a rectangular region with dimensions $L_x$, $L_y$, and $L_z$. The structure is illuminated by a plane wave propagating in the xz-plane with the incident angle of $\theta_i$. The area between the dielectric particles is filled with air and the scattering is calculated in the upper half plane for $-\pi/2 \leq \theta_s \leq \pi/2$.

To calculate the scattered field, we solve for the total scalar field amplitude $U(\mathbf{r})$ satisfying the general Lippmann-Schwinger integral equation given in Eq. (1) and independently solved for each realization. The total field amplitude at $\mathbf{r}$ can be constructed from the incident field amplitude $U_i(\mathbf{r})$ and the contributions from all scatterers, where $G_0(\mathbf{r},\mathbf{r}')$ is the background Green's function [39] and $k_0$ is the free space wavenumber

$$U(\mathbf{r}) = U_i(\mathbf{r}) + k_0^2 \int d^3\mathbf{r}' G_0(\mathbf{r},\mathbf{r}') \eta(\mathbf{r}') U(\mathbf{r}'). \qquad (1)$$

The susceptibility parameter $\eta(\mathbf{r})$ is defined such that it is locally related to the relative permittivity by $\varepsilon_r(\mathbf{r}) = 1 + 4\pi\eta(\mathbf{r})$ [39]. For a collection of discrete point-like scatterers (i.e., delta function approximation), $\eta(\mathbf{r})$ may be described as a summation over the effective polarizability of particles defined at the center of each particle as $\eta(\mathbf{r}) = \sum_{i=1}^{N_p} \alpha_0 \delta(\mathbf{r} - \mathbf{r}_i)$, in which $\alpha_0 = r^3(n^2 - 1)/3$ [40]. The free space Green's function $G_0(\mathbf{r},\mathbf{r}') = e^{ik_0|\mathbf{r}-\mathbf{r}'|}/|\mathbf{r}-\mathbf{r}'|$ is utilized, following the $e^{-i\omega t}$ convention, with proper adjustments for the regularized Green's function at $\mathbf{r} = \mathbf{r}'$ [41]. In the final step, the scattering amplitude of the outgoing spherical wave, in the far field domain, is extracted and studied. We model $N_p$ identical dielectric particles (i.e., simulating commonly used $SiO_2$ particles with refractive index $n = 1.5$) and subwavelength radius of $r = \lambda_0/2\pi$. Clearly, the most accurate solution may be attained by using the complete Mie scattering theory for a composition of spheres while solving for the fields using the exact Maxwell's equations [32,42]). The computation time and complexity of such analysis, however, drastically increases with increasing the number of the scatterers and the size of the simulation domain. On the other hand, the scalar approximation [1,6,17,38] enables us to capture many important underlying physical properties of the system, while avoiding unrealistically long simulation times.

Several physical factors must be considered in the choice of the material and the size of the scattering bodies. In particular, if the scattering elements are extremely off resonance (i.e., with extinction cross sections much smaller than the physical cross section of the particles), one might expect that millions of particles are required to create a meaningful scattering response, associated with large mean free path lengths [43].

Along the same line of physical reasoning, we expect to observe meaningful coherent scattering responses and weak localization effects from a smaller number of scatterers when the scattering amplitude is higher. Using Mie theory, the extinction efficiency (i.e., extinction cross section normalized to the physical cross section of the particle) for the dielectric spheres (scattering centers) modeled in this work is found to be 0.22, the amplitude of the electric dipole scattering coefficient is 0.19, and the amplitudes for all other higher order scattering coefficients are smaller than 0.03. We note that the response of each single particle is still well within the off-resonant dipolar approximation. For the lossless configurations examined here the extinction cross section is equal to the scattering cross section and represents the amount of power that the object extracts from the input wave normalized to the incident power intensity.

In the following, we study the formation of the CBS cone from various finite-sized media based on the above formulation. In addition to the conventional parameters applicable in studying semi-infinite random media, such as density factor and angle of incidence, here we also study parameters specifically relevant to finite-sized systems including the absolute number of scatterers and the size of the domain. The ensemble-averaged specular reflection in multiple scattering and a technique to eliminate it is also discussed and the modified results are presented.

## 3. Results

We analyze several examples of random geometries (such as that depicted in Fig. 1) following the numerical formulation described in the previous section. In all the examples, the random domain is a cube, i.e., $L_x = L_y = L_z = L$ and density factor is the ratio of the volume of all particles to the volume of the cube, defined as

$$\rho = N_p \frac{V_S}{L^3}, \tag{2}$$

in which $V_s$ is the volume of each dielectric sphere. The $\rho$ parameter is an indication of how closely the particles are packed. The scatterers are assumed to consist of a low-index material ($n=1.5$) with a radius of $r=\lambda_0/2\pi$. As discussed above, the Mie scattering analysis of individual particles shows that their response lies well within the single-mode dipolar scattering regime and higher order scattering modes are negligible. For all the results provided in Fig. 2 to Fig. 7, $N_r$ distinct random configurations are simulated, and the results are averaged to reveal the coherent contributions vs. the incoherent background signals.

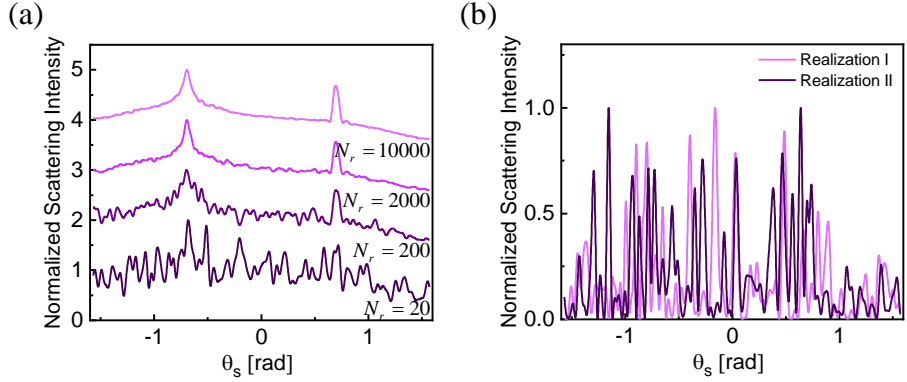

Fig. 2. (a) Normalized ensemble-averaged scattering intensity for configurations with $N_p=10000$, $L=20\lambda_0$, $\theta_i=40$ [deg] $\approx 0.7$ [rad], and $\rho=0.021$. Results are averaged over $N_r=(20,200,2000,10000)$ distinct realizations of the disorder, with darker colors corresponding to smaller $N_r$. The maximum of all curves is normalized to two and for the sake of better visualization, each curve is moved one unit up compared to the previous one. (b) Angular dependance of scattered intensities (speckle patterns) for two sample realizations.

We first assess the effect of disorder averaging. For samples with a very large number of particles (the case typical of most laboratory experiments), usually tens to hundreds of measurements are sufficient for the speckle to average to a diffuse background allowing the CBS cone to emerge [44]. This relatively low number of realizations is not sufficient to observe the weak localization effects in small, disordered geometries. By increasing the number of realizations, however, the localization effects appear for these geometries as well. This important conclusion is carefully examined here. Keeping all the other parameters unchanged, Fig. 2(a) illustrates the evolution of the averaged scattering profile when $N_r$ is increased from 20 to 10000 in a random sample with $N_p=10000$, $L=20\lambda_0$, $\theta_i=40$ [deg] $\approx 0.7$ [rad]. This corresponds to a density factor of $\rho=0.021$ or approximately 2% volume fraction. Here, $\lambda_0$ indicates the free space wavelength.

First, we notice that ensemble averaging over 2000 (or even smaller) number of simulations is sufficient to capture the coherent scattering response in these geometries. The CBS cone is still captured for averaging over 200 samples, however as $N_r$ increases the response across all angles becomes less noisy. Another important observation is that the ensemble-averaged specular coherent reflection is present at $\theta_s=\theta_i$ in averaging over 200 realizations or more This effect is well separated from the CBS contribution under oblique incidence and no specular reflection is present in scattering from any single realization (Fig 2(b)) or small values of $N_r$. Note that the background material here is air everywhere; thus, all the observed specular effects are the direct consequence of embedded coherence effects in multiple scattering, outliving the ensemble averaging step. The CBS ratio is slightly smaller than the theoretical value of two and does not change noticeably by increasing the number of realizations. The finite size of the scattering medium plays a fundamental role in establishing the width of the CBS cone. For finite-sized samples, such as the examined cases here, the probability of having longer scattering trajectories for photons severely diminishes, and thus the width of the CBS cone is anticipated to increase [45,46,47]. This is consistent with our observations here (see Fig. 2 to Fig. 7) where the finiteness of the samples results in broader CBS cones.

Performing a statistical analysis over multiple implementations of the random medium provides insight into the coherent wave effects inside the random structure; at the same time, it is also insightful to investigate the scattering response for individual realizations. Here, to emphasize the effect of ensemble averaging, the angular distribution of scattering intensity for two sample realizations (i.e., $N_r = 1$) is shown in Fig. 2(b). These curves demonstrate the speckle patterns in the far field domain, originating from interference of waves due to multiple scattering paths.

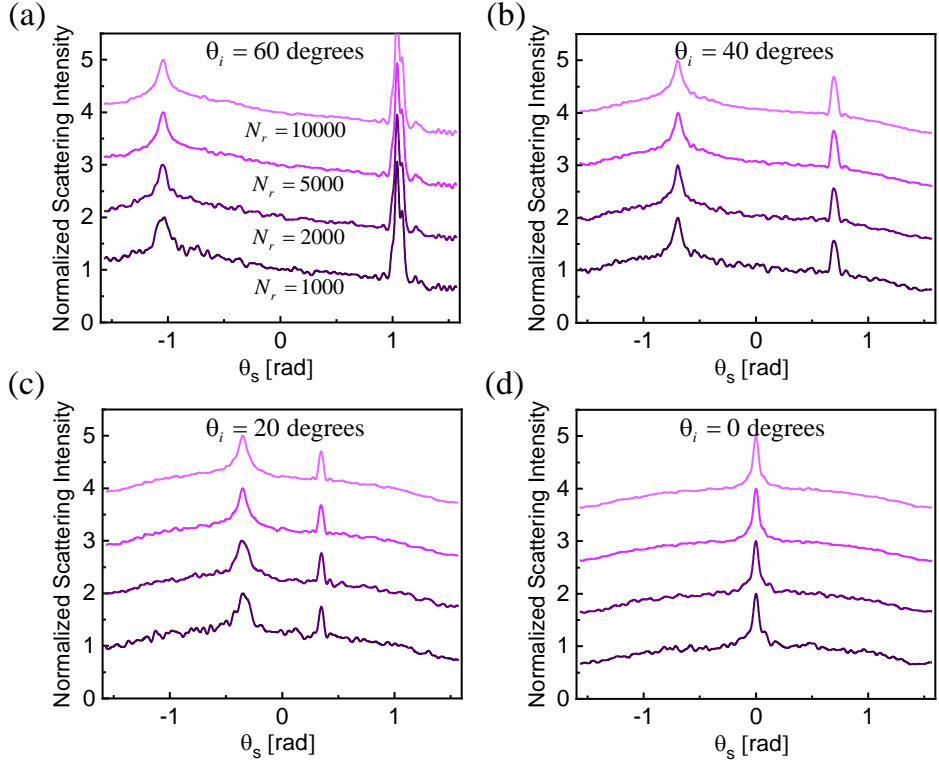

Fig. 3. Normalized ensemble-averaged scattering intensity for random samples with $N_p = 10000$, $L = 20\lambda_0$, and $\rho = 0.021$. Results are averaged over $N_r = (1000, 2000, 5000, 10000)$ different simulations, with darker colors corresponding to smaller $N_r$ in each case. The maximum averaged intensity in each curve is normalized to two, and for better visualization each curve is moved one unit up compared to the previous one. The angle of incidence is set at (a) $\theta_i = 60$ [deg], (b) $\theta_i = 40$ [deg], (c) $\theta_i = 20$ [deg], and (d) $\theta_i = 0$ [deg].

Next, we investigate the effect of the angle of incidence on the emergence of weak localization and the degree of enhancement as well as the evolution of the specular fraction of the coherent scattering. A particularly interesting case is when $\theta_i = 0$ [deg], in which the backscattering and specular directions overlap. This is shown in Fig. 3 where we examine four incident angles, $\theta_i = (60, 40, 20, 0)$ [deg] covering from near-grazing angles to exact normal incidence. Physical properties of the sample are unchanged compared to the results in Fig. 2 and ensemble averaging over $N_r = (1000, 2000, 5000, 10000)$ simulations are performed for each oblique/normal illumination.

Examining the results in Fig. 3, the width of the CBS cone and the enhancement ratio is approximately independent of the angle of the incidence except for the case of normal incidence. The fact that the width of the peak only depends on the properties of the sample is consistent with previous studies indicating a direct relation between the width of the CBS cone and the mean free path length inside the disordered medium [1,14,48]. We also note that the width of the cone may be slightly overestimated for smaller values of $N_r$. Interestingly, for the case of normal incidence, we notice that both the line shape and width of the CBS cones deviate from the results for oblique incidence. This is due to the superimposition of the specular reflection and the coherent backscattering at this angle, which makes it difficult to accurately estimate the line shape and width of the CBS cones.

Indeed, an interesting observation in all the numerical results presented so far is the occurrence of specular reflection arising in scattering from the random samples. Experimentally, to avoid the specular reflection, random samples are typically tilted off axis and ensemble averaging over these angles are not of interest [6]. However, here, we closely examine the evolution of averaged intensity over these angles and devise a technique to explicitly eliminate averaged specular peaks while maintaining the localization contributions unchanged. Numerical results in Fig. 3 also indicate that quite contrary to the CBS, specular intensities are very sensitive to the angle of incidence. For near-grazing angles, both the intensity and width of the specular reflection are more pronounced compared to normal illumination.

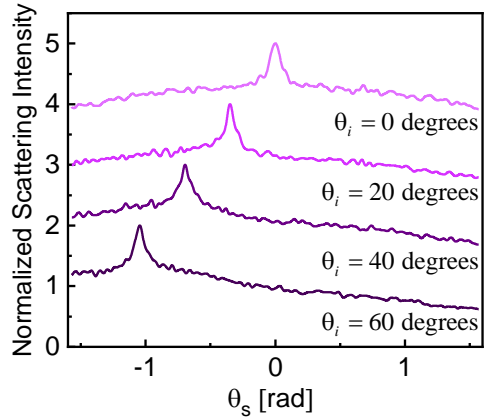

Fig. 4. Elimination of specular reflection from ensemble-averaged scattered intensity. Random samples are similar with those used in Fig. 3, where an artificial random surface is added on top of the medium shown in Fig. 1. The angle of incidence is set at $\theta_i = (60, 40, 20, 0)$ [deg] (different colors). The darker colors corresponding to higher incidence angles. $N_r = 2000$ and the maximum value (CBS peak) of all curves are normalized to two. Each curve is moved one unit up compared to the previous one for better visualization.

The emergence of the specular reflection, emerging from statistical averaging, can be interpreted as the direct consequence of the "artificial interface" present at $z = L_z$. This artificial interface is built by restricting the z position of all particles to be inside the cube. For finite-sized samples, as is the case here, we observe comparable specular and CBS intensities. To eliminate the specular peak, the artificial interface (on top of the block) is replaced by an effective "random surface" stretching over a finite thickness around $z = L_z$. The local position of the artificial random surface at each point in space is determined using semi-Fourier series functions constructed using trigonometric functions. As the result, the new random medium is uneven at the top interface and the coherent properties at the specular angle are expected to diminish. This is indeed the case as shown in Fig. 4, for random samples with $N_p = 10000$, $L = 20\lambda_0$, and $N_r = 2000$ examined under various illumination angles. Compared with Fig. 3, the specular wave is clearly suppressed and is totally undetectable. Notably, the CBS cone under the normal incidence is now wider (comparing Fig. 3(d) and Fig. 4) and the observed cone width is consistent with the widths under oblique illumination as well. Also, the degree of enhancement for normal incident wave modifies to a value slightly below 2 in the new simulation, consistent with the ratios observed at oblique incidence, as expected. Therefore, this approach makes it possible to accurately characterize the coherent backscattering effect, even at normal incident angles.

We next examine the line shape of the CBS cones observed for the finite-sized, wavelength-scale geometries, and compare these intensity profiles with theoretical predictions traditionally made assuming large configurations. The angular scattering response of the medium contains information about multiple scattering phenomena and the scattering path lengths. Accordingly, here we utilize the numerically calculated intensity angular profiles to extract information about the mean free path length. In the case of a semi-infinite scattering medium, the backscattering cone follows a triangular shape (i.e., linear behavior near the backscattering direction). When the medium is of finite size or finite thickness, the triangular line shape becomes rounded due to the omission of the higher-order multiple scatterings [1, 45]. Generally, the line shape of the backscattering cone depends on the wavelength and the mean free path length inside the random media [1, 5, 43]. To this end, we compare the profile of the localization here with the theoretical model devised by Akkermans and Montambaux [43] with the correction for finite-sized random domains

$$\alpha_d(L) = \frac{3}{8\pi}\left(1 - e^{-2b}\right)\left(1 - \frac{\tanh(b/2)}{b/2}\right)$$
$$\alpha_c(k_\perp, L) = \frac{3}{8\pi}\frac{1 - e^{-2b}}{(1 - k_\perp l_e)^2}\left[1 + \frac{2k_\perp l_e}{(1 + k_\perp l_e)^2}\frac{1 - \cosh(b(1 + k_\perp l_e))}{\sinh(b)\sinh(bk_\perp l_e)}\right]. \tag{3}$$

Here, $b = L/l_e$, where $L$ is the thickness of the sample and $l_e$ is the mean free path length used here as a fitting parameter. $k_\perp$ is the amplitude of the transverse projection of the vector $\mathbf{k}_i + \mathbf{k}_s$ on the xy plane, in which $\mathbf{k}_i$ and $\mathbf{k}_s$ are incident and scattering wavevectors, respectively [43].

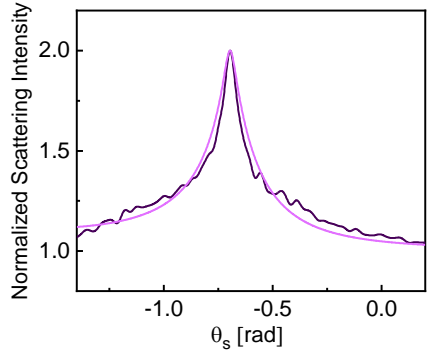

Fig. 5. Numerically calculated normalized ensemble-averaged scattering intensity for $N_p = 10000$, $L = 20\lambda_0$, and $\rho = 0.021$ for $N_r = 10000$ (dark curve) and theoretical fit (Eq. (3)) with $l_e$ used as fitting parameter (light color). Note that $l_e$ value is normalized to the free-space wavelength.

Figure 5 compares the numerically calculated line shape with a theoretical fit using Eq. (3), where $l_e$ is estimated at $0.94\lambda_0$ for a sample with 10000 particles and box size of $L = 20\lambda_0$. First, we note that the shape of the backscattering cone for this finite, wavelength-scale geometry using the scalar wave analysis closely follows the analytical predictions. In addition, the thickness of the box is much larger than the derived mean free path length in this example, justifying the validity of employing these theoretical predictions to model such small samples [43]. The multiple scattering is happening here in the weakly disordered regime as $kl_e \approx 5.9 > 1$, slightly above the Ioffe-Regel criterion [49] of $kl_e \simeq 1$ for the appearance of Anderson localization. The CBS cone width is inversely proportional to the mean free path length and may be estimated using the formula $FWHM \approx \lambda_0 / 3\pi l_e$ [14,48]. Utilizing this formula, the mean free path length in this example is approximated to be $l_e = 0.66\lambda_0$, in reasonable agreement with the value derived from the theoretical fit above.

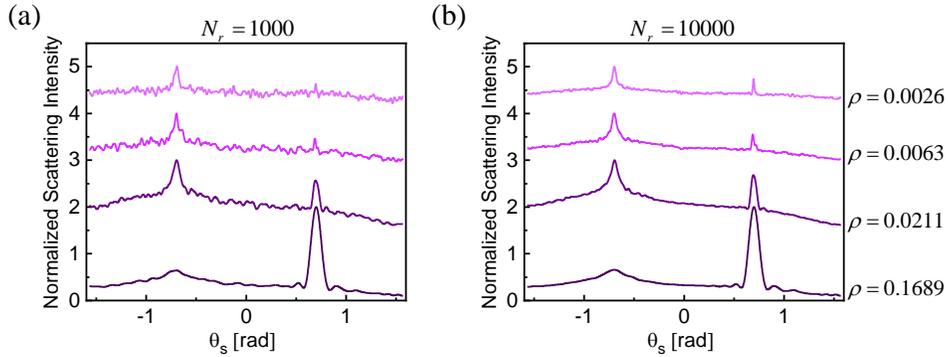

Fig. 6. Normalized ensemble-averaged scattering intensity for random samples composed of $N_p = 10000$ particles, where the size of the random media is increased from $L = 10\lambda_0$ to $L = 40\lambda_0$ in equal steps of $10\lambda_0$ (darker colors correspond to smaller boxes). Angle of incidence is fixed at $\theta_i = 40$ [deg] in all cases and the maximum of all curves is normalized to two. For clarity, each curved is moved one unit up compared to the previous one. The ensemble averaging is performed over (a) $N_r = 1000$ and (b) $N_r = 10000$ distinct realizations.

To better understand the effects of sample size, number of particles, and density factor on the shape of the cone, in the following we investigate two specific cases where we fix the density factor while changing the box size (Fig. 6), and number of particles (Fig. 7).

Figure 6 illustrates the combined effects of density factor and sample size on the characteristics of the CBS cone. The number of particles is set at a reasonably high value of 10000, while the box size is gradually increased from $10\lambda_0$ to $40\lambda_0$. This is associated with decreasing the density factor from approximately 17% to 0.26% and we conduct the simulations for 1000 (Fig. 6(a)) and 10000 (Fig. 6(b)) independent realizations. Figure 7 shows the combined effects of the absolute number of particles (i.e., $N_p$) and sample size, for two fixed density factors. In Fig. 7(a), density factor is fixed at 0.63% and the box size is gradually increased from $L=10\lambda_0$ to $L=34.3\lambda_0$, corresponding to samples with very few particles (i.e., $N_p = 370$) to very large number of particle numbers (i.e., $N_p = 15000$), respectively. In Fig. 7(b), the density factor is set at 2% and the box size is changed from $L=10\lambda_0$ to $L=25.2\lambda_0$.

Several interesting observations can be made from the results presented in Fig. 6 and Fig. 7. First, samples with lower density factor are more sensitive to ensemble averaging, so a higher number of independent realizations is required to retrieve the embedded coherent effects. In addition, the width of CBS cone is slightly overestimated for smaller $N_r$, also more evident for sparse samples. Second, the intensity of the specular reflection is directly related to the density factor, as expected. At the studied size range (between $L=10\lambda_0$ to $L=34.3\lambda_0$) the width of the specular reflection is slightly dependent on the size of the sample, and we observe narrower specular reflections for larger boxes. This width, however, is significantly decreased for sparse samples (Fig 6). Third, we notice that the coherent contributions in the backscattering still emerge for even as few as a few hundred particles (Fig. 7(a)). To capture such coherent features, ensemble averaging on many different samples may be required. Fourth, the width of the CBS cone is strongly proportional to the density factor with a slight inverse dependence on the absolute number of particles for these samples. This is consistent with previous studies on the role of mean free path length on the shape of the cone, $FWHM \propto 1/l_e$ [14,48], indicating longer path lengths for more sparse samples as long as the sample size is much larger than $l_e$. Fifth, the enhancement ratio is closer to the theoretical limit of two for samples with higher density factor, consistent with previous analysis of electromagnetically large random samples [5,27,46]. In the case of constant density factor, increasing the sample volume (i.e., more particles) also significantly increases the enhancement ratio (Fig. 7).

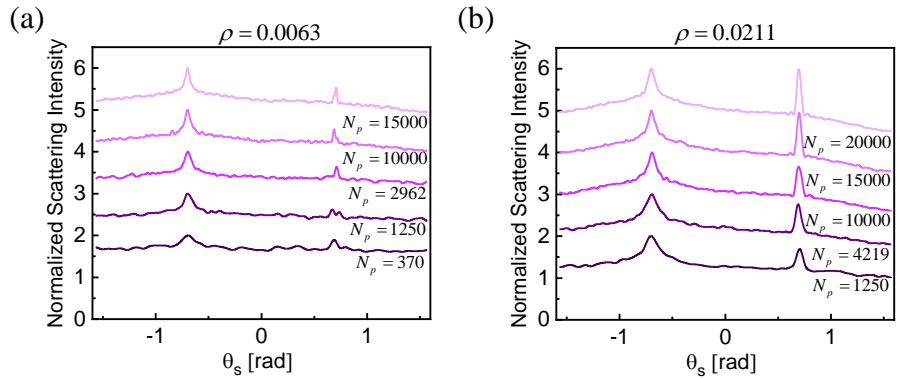

Fig. 7. Normalized ensemble-averaged scattering intensity for two different density factors $\rho$ while the number of particles and the size of the enclosing box are changed. Curves correspond

to (a) $\rho = 0.0063$, $L = (10, 15, 20, 30, 34.3)\lambda_0$, and (b) $\rho = 0.021$, $L = (10, 15, 20, 22.9, 25.2)\lambda_0$ with darker colors corresponding to smaller boxes and smaller number of particles. Ensemble averaging is performed over $N_r = 5000$ realizations and the angle of incidence is fixed at $\theta_i = 40$ [deg]. In all cases the maximum value is normalized to two and for clarity, each curved is moved one unit up compared to the previous one.

## 4. Conclusions

In this work we have provided a comprehensive study of the coherent backscattering from finite-sized sparse scattering medium based on a scalar wave analysis. CBS is typically associated with semi-infinite samples with very large number of scatterers. Relying on a simplified form of Maxwell's equations in a multi-scattering medium, we investigated the effect of density factor, number of particles, and size of the scattering medium on different characteristics of the CBS line shape. The density factor was shown to be the prominent factor in determining the width of CBS cone, while both the density factor and the number of particles influence the degree of enhancement in the backscattering direction. The angular distribution and the enhancement factor of the scattering intensity hold valuable information to characterize random samples. The cone shapes are found to be in very good agreement with previous theoretical calculations. We have also extracted the mean free path length of the wave inside the random media from the angular dependence of the backscattered wave in our simulations. Finally, we have devised a technique to exclusively eliminate the specular reflection component from the ensemble-averaged backscattered intensities, enabling the observation of pure CBS line shapes even for normal illuminations of the structure. This work provides physical insight into the multiple scattering of photons and can serve as the basis for studying more complex incident fields in characterization of multiple scattering phenomena.

**Disclosures.** The authors declare no conflicts of interest.

**Acknowledgements.** The authors acknowledge fruitful discussions with John Schotland.